# Self-regulated charge transfer and band tilt in nm-scale polar GaN films

M. –H. Tsai [*a], and S. K. Dey[b]

[a] *Department of Physics, National Sun Yat-Sen University, Kaohsiung 80424, Taiwan*

[b] *Department of Chemical and Materials Engineering, & Electrical Engineering, Ira A. Fulton College of Engineering, Arizona State University, Tempe, AZ 85287-6006, USA*

To date, the generic polarization of Bernardini, Fiorentini and Vanderbilt ($P_{BFV}$) has been widely used to address the issue of polarity in III-V nitride semiconductors, but improvements in nitride materials and the performance of optoelectronic devices have been limited. The current first-principles calculation for the electronic structures of nm-scale [0001] GaN films show that the internal electric fields and the band tilt of these films are in opposite direction to those predicted by $P_{BFV}$. Additionally, it is determined that an intrinsic self-regulated charge transfer across the film limits the electrostatic potential difference across the film, which renders the local conduction band energy minimum (at the Ga-terminated surface) approximately equal to the local valence band energy maximum (at the N-terminated surface). This effect is found to occur in films thicker than ~4nm.

PACS numbers: 73.61.Ey, 73.90.+f

## 1. Introduction

Wide-band-gap semiconductors, including III-V nitride semiconductors, are important materials for optoelectronics applications that have been intensively investigated [1-25]. The polarity of III-V semiconductors is a critical issue that has hindered improvements in nitride materials and the performance of optoelectronic devices [13]. In the wurtzite polymorph, these materials are usually grown along the c-axis. Both the incorporation of impurities and the formation of defects are related to the polarity that depends on the growth mechanism. Moreover, the occurrence of a two-dimensional (2-D) electron gas and observed optical properties of heterostructures are influenced by internal field effects caused by the polarization (associated with the polarity). Bernardini, Fiorentini, and Vanderbilt (BFV) have calculated the polarizations of various wide-band-gap III-V semiconductors including GaN [15]. The calculated polarizations ($P_{BFV}$'s) by BFV, in conjunction with piezo-electric polarizations [16], have been widely applied to determine polarity, tilt of energy bands or band profile, and sheet concentration of 2-D electron gas at interfaces [14]. Hellmann [12] and Sumiya and Fuke [13] have published critical reviews on the determination and control of polarity as well as "spontaneous and piezo-electric" polarizations.

BFV used the bulk model and obtained structural parameters for hexagonal GaN, i.e., $a=6.040 a_o$ (where $a_o$ is the Bohr radius), $c/a=1.6336$, and $u=0.376$ [15]. The bulk model used by solid-state-physics theorists has a periodic boundary condition, which is equivalent to an infinitely repeated boundary-free (or surface-free) imaginary solid. This model can be viewed as $GaN_4$ tetrahedrons, arranged in a hexagonal array, with apex pointing in the $+c$ (or [0001]) direction. Since each N ion at the corner of the tetrahedron is shared by four tetrahedrons, the charge partition in each tetrahedron is $Ga^{+Z}N_4^{-Z/4}$ if one assumes that the effective charge, Z, in the N ion can be equally divided. If $u_{ideal}=3/8=0.375$, each $Ga^+$ ion would coincide with the negative charge centroid of four surrounding $N^-$ ions and there will be no polarization as in the zinc-blende polymorph. However, since $u$ is slightly larger than 0.375 in the case of GaN, there is a small local dipole in the $-c$ direction within a tetrahedron due to the shift of the negative charge centroid in the $+c$ direction. Using the simplified point charge model with the $Ga^{+Z}N_4^{-Z/4}$ partition and an effective charge (Z=2.72e) obtained by BFV [15] and $u=0.3764$, the dipole moment density, i.e. the polarization, is $P=-(u-0.375)cZ/V=-0.014 C/m^2$; where V is the volume of the basic $GaN_4$ tetrahedron. This value is about half of $P_{BFV}$ ($-0.029\ C/m^2$). The same order of magnitude and sign suggest that the origin of $P_{BFV}$ may simply stem from the deviation of $u$ from $u_{ideal}$, as supported by their calculated trend: $u$'s of 0.376 (GaN), 0.377 (InN), and 0.380 (AlN) gives $P_{BFV}$'s of $-0.029$, $-0.032$, and $-0.081\ C/m^2$, respectively [15].

For films, the $Ga^+$ ions on the Ga-terminated surface don't belong to any complete $GaN_4$ tetrahedron, so that the $+Z$ charge on each Ga surface ion is not counted in $P_{BFV}$. But, each of the 2nd-layer N ions belongs to one $GaN_4$ tetrahedron, so that $-Z/4$ of the N-ion charge is included in $P_{BFV}$ and $-3Z/4$ is not counted. Totally, the Ga-terminated surface has an uncounted 2-D charge density of $+0.022 e/a_o^2$. On the other side of the film, each N ion on the N-terminated surface belongs to three complete tetrahedrons, so that $-3Z/4$ of the N-ion charge is included in $P_{BFV}$ and $-Z/4$ is not counted. Thus, this surface has an uncounted 2-D charge density of $-0.022 e/a_o^2$. If the $Ga_4N$ tetrahedron is regarded as the basic unit instead, the uncounted 2-D surface charge densities are still the same. In comparison, the 2-D charge densities of $-\sigma_{pol}$ and $+\sigma_{pol}$ ($\sigma_{pol}=|P_{BFV}|=0.00051 e/a_o^2$) due to the termination of $P_{BFV}$ at the Ga- and N-terminated surfaces, respectively, are negligibly small and in opposite sign.

The bulk model and the corresponding symmetry have been successful for describing interior properties of low-ionicity materials and nonpolar films of ionic materials. However, for polar films of high-ionicity materials such as [0001] GaN films, the long ranged nature of the electrostatic potential dictates that the electric field due to surface or interface charges still exists even in the interior of a thick film like a capacitor, in which the symmetry of an infinitely extended, imaginary solid no longer exists. This argument shows that the use of $P_{BFV}$, which is based on the bulk symmetry, to represent the electrostatics in these films is not adequate, let alone the neglect of the dominant uncounted surface charges. Thus, a theoretical study based on a more realistic film model and a new perception of the ionic charge arrangement is needed.

## 2. A new perception of the electrostatics in polar GaN films

Since the c-axis oriented GaN film is composed of an array of $Ga^+$-$N^-$ (or $N^-$-$Ga^+$) bi-layers and each bi-layer is a dipole layer, it is more natural to perceive the film as an array of dipole layers. The accumulation of the electrostatic potential differences along the dipolar array can give rise to a tilt in the energy bands of a film. Therefore, when a film is thick enough, the local conduction band minimum (CBM) at the Ga-terminated surface, $S_{Ga}$, becomes lower than the local valence band maximum (VBM) at the N-terminated surface, $S_N$, as shown in Fig. 1. Since this is no longer a stable electronic system, a charge transfer of electrons from occupied states near VBM at $S_N$ to empty states near CBM at $S_{Ga}$ must occur. As a consequence of this charge transfer across the film, the resulting band diagram may be represented as in Fig. 2. Fiorentini et al. [18] envisioned a band-gap limited polarization field in thick films by photo or thermally excited free carriers, which is completely different from those discussed here. First, the field envisioned by Fiorentini et al. is in opposite direction. Second, the charge transfer effect discussed here occurs even at zero temperature, nm-scale thickness and without any excitation. Third, the resultant field envisioned by them is constant across the film, which gives rise to a linear potential profile different from that depicted in Fig. 2.

If $V_{bi-layer}$ is the electrostatic potential difference across a $Ga^+$-$N^-$ bi-layer, the critical number of bi-layers, $N_C$, for the onset of VBM($S_N$)→CBM($S_{Ga}$) charge transfer is $N_C \sim E_g/qV_{bi-layer}$, where $E_g$ is the band gap and q is the electronic charge. Thus, when the critical thickness of a film is exceeded, i.e., when $N > N_C$, the ensuing charge transfer will limit the electrostatic potential energy difference between $S_{Ga}$ and $S_N$ to be a constant, and would equal the sum of $E_g$ and the potential energy differences at both $S_{Ga}$ and $S_N$ due to surface dipole layers associated with dangling bonds or surface states. For multi-layer quantum wells, the surface contribution should be replaced by interface contributions.

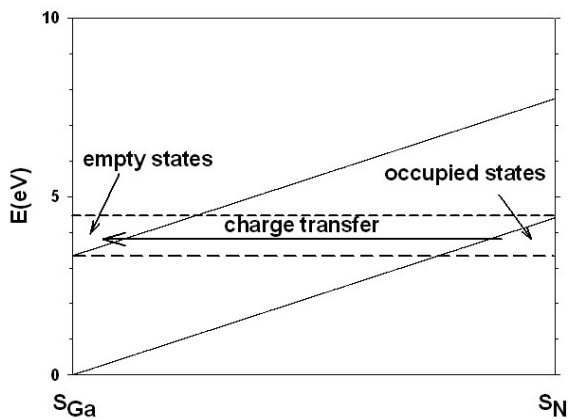

Fig. Band tilt due to accumulation of electrostatic potential differences across dipolar bi-layers.

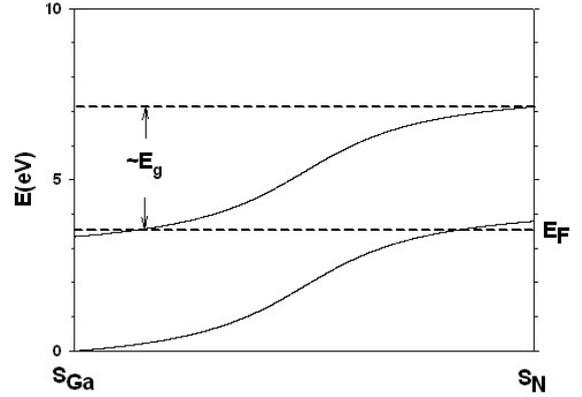

Fig. 2. Self–limiting tilt of energy bands due to charge transfer across the film. Upper and lower solid lines show the local conduction band minimum (CBM) and local valence band minimum (VBM), respectively.

## 3. Calculation method

The first-principles calculation method used in this study is the pseudofunction (PSF) method [26]. The simple mixing scheme with an extremely small mixing factor of 0.0002 or 0.0001 was used for self-consistent iterations due to instability of the subtle charge transfer across the film. The Bloch sums of auxiliary PSFs and charge density/potential are expanded in 10043 and 72765 plane waves, respectively, for the 12-bi-layer thick film. In the calculation of the self-consistent potentials, the six special-$k$ point scheme of Cunningham for hexagonal two-dimensional lattice [27] is used to approximate the integration over the first Brillouin zone.

The experimental lattice constants of GaN of $a$=3.190Å and $c$=5.189Å are employed and $u$ is chosen to be 0.3766. The Ga-terminated [0001] GaN surface has various surface reconstructions, whereas the N-terminated [000$\underline{1}$] GaN surface relaxes to the unreconstructed (1x1) surface structure [28]. A different surface structure gives rise to a different surface contribution, which only causes a rigid shift of the electrostatic potential across the film. Since the central theme of this study is to demonstrate the effect of self-limitation of the electrostatic potential across the film and the consequent tilt of the energy bands, thin films with unreconstructed and unrelaxed surfaces should suffice to meet our objective. Thus, in this study, 2- to 12-bilayer thick films with only unreconstructed (1x1) and unrelaxed surfaces have been considered.

## 4. Calculation results and discussion

For wurtzite films with an even number of bi-layers, the (x,y) coordinates of the ions in the top $S_{Ga}$ surface layer are the same as those in the bottom $S_N$ surface layer. Thus, only films with an even number of bi-layers, which have a common surface contribution to the potential difference across the film, have been considered. Figiure 3 shows the potential energy difference, q$\Delta$V, between $S_{Ga}$ and $S_N$ as a function of the thickness, $t$ (~0.26 nm/bilayer). The curve shows that q$\Delta$V approaches a constant (saturated) value, within 12 bi-layers, as $t$ is increased. The limiting q$\Delta$V is about 4.22 eV, which is the sum of $E_g$ and potential energy differences due to surface dipole layers associated with dangling bonds or surface states. The saturation of q$\Delta$V clearly shows the self-regulated electron charge transfer from the $S_N$ side to the $S_{Ga}$ side that maintains a constant electrostatic potential difference ($\Delta$V) across the film.

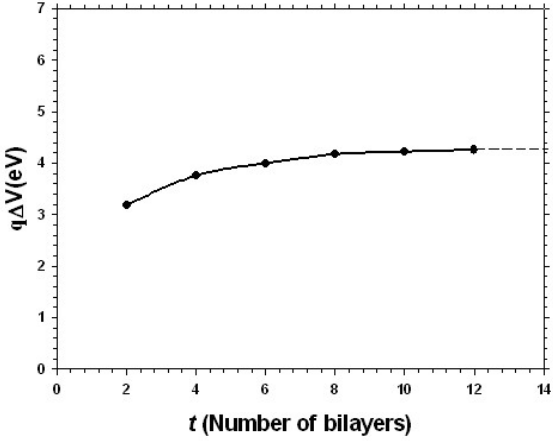

Fig. 3. The potential energy difference, q$\Delta$V, across the film as a function of the thickness, $t$, given as number of bi-layers for [0001] Wurtzite GaN. The filled circles represent calculated values.

Figure 4 shows the local VBM (denoted by open circles) and CBM (denoted by open triangles) as a function of the distance, z, from $S_{Ga}$ in the 12-bi-layer thick film. The local VBM (CBM) is the average energy of the $\Gamma$-point states, which contain significant charges inside the muffin-tin sphere of the $N^-$ ($Ga^+$) ions located at z. The average separation in energy (or $E_g$) between the local CBM and VBM in the whole range of z is about 2.0 eV. This is significantly smaller than the experimentally obtained $E_g$ of GaN (3.34 eV at 300K [29]); the underestimate of $E_g$ is a well-known deficiency of the local density approximation for the exchange-correlation potential.

The calculated band bending or tilt shown in Fig. 4 is similar to the schematic of Fig. 2. The dashed line indicates the approximate equality of the local CBM near $S_{Ga}$ and the local VBM near $S_N$. The Fermi level, $E_F$, depends on the dangling-bond/surface states, so that it differs from that shown in Fig. 2.

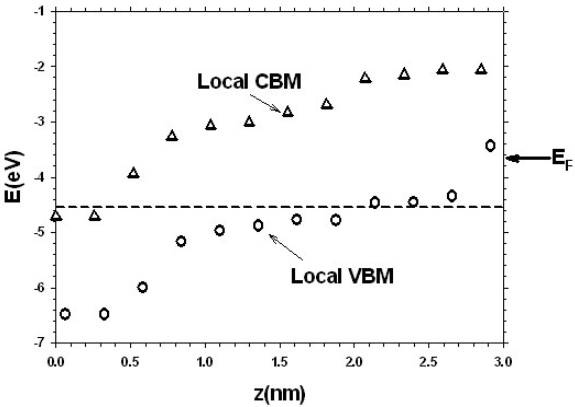

Fig. 4. The local CBM and VBM as a function of the distance, z, form $S_{Ga}$ in 12-bi-layer thick (0001) Wurtzite GaN. The dashed line corresponds to the $E_F$ shown in Fig. 2, which shows the line up of the local CBM near $S_{Ga}$ and the local VBM near $S_N$. Note that the actual $E_F$ depends on the dangling-bond/surface states.

The critical number of bi-layers, $N_C$, discussed previously depends on $E_g$. The saturated potential energy difference, q$\Delta$V, across the film also depends on $E_g$. Since $E_g$ is underestimated by 40% of the experimental value, $N_C$ and saturated q$\Delta$V are also underestimated. The accuracy of $E_g$ can be improved by replacing the exchange part of the exchange-correlation energy by the exact exchange energy of Görling [30]. However, the contribution of $E_g$ to $N_C$ and saturated q$\Delta$V can be expected to be linear. Then, the present study suggests that the correct $N_C$ be within ~17 bi-layers (or ~4 nm) and the saturated q$\Delta$V be ~5.6eV. For multi-layer quantum wells, the surface contributions to the saturated q$\Delta$V should be replaced by interface dipolar contributions.

In this study, the maximum thickness considered is only 12 bi-layers or ~3 nm. The microscopic pictures (Figs. 4) of the band tilt or band bending are not as smooth as the macroscopic, averaged picture depicted in Fig. 2. However, the self-regulated charge transfer effect and, therefore, the tilt of energy bands between $S_{Ga}$ and $S_N$ may be expected in thicker [0001] GaN films, zinc-blende (111) GaN films and the polar films of other ionic semiconductors. The internal electric field, $E_{int}$, which is the negative of the slope of the electric potential, and the band tilt obtained in this study are in opposite direction to those predicted by $P_{BFV}$. This discrepancy apparently is due to the neglect of the uncounted $\pm 0.022e/a_o^2$ surface charges in using $P_{BFV}$. Besides, the present study implies that $E_{int}$ decreases with the increase of the film thickness. ($E_{int}$ is roughly equal to $E_g/qt$.) In contrast, $E_{int}$ is independenent of the film thickness according to $P_{BFV}$, if charge transfer is ignored.

Since Ambacher et al.[14] claimed experimental evidence of $P_{BFV}$, it is necessary to clarify this issue. Ambacher et al.[14] used $P_{BFV}$ and piezo-electric polarization to determine the sheet concentration of a 2-D electron gas at the AlGaN/GaN interface, which was then correlated with the sheet concentration obtained by Hall measurements to provide evidence of $P_{BFV}$. The sheet concentration calculated by Ambacher et al. is the concentration of the conduction electrons, which screen the 2-D charge ($\sigma_{pol}$) arising from the discontinuity of $P_{BFV}$ at the AlGaN/GaN interface. However, Ambacher et al.'s calculation model ignored the following crucial effects that dramatically influence the band profile. First, Ambacher et al. [14] ignored the charge transfer effect due to the band offset or the line-up of chemical potentials in AlGaN and GaN layers; the difference in the band gap between $Al_xGa_{1-x}N$ (0.24<x<0.31 for sample A [14]) and GaN is ~0.74eV, which is not negligible. Second, the long-range effects of the screening charges in the electrodes and the $\sigma_{pol}$'s at the electrode/AlGaN, GaN/AlGaN and AlGaN/substrate interfaces were all ignored [14]. Third, with a static permittivity of $\varepsilon$ of 9.04 [31], the electric field arising from the termination of $P_{BFV}$ at both ends of the GaN layer (i.e., $\pm\sigma_{pol}=\pm 0.00051e/a_o^2$), is $3.3\times10^9/\varepsilon$ V/m without a band-gap limiting effect; translating to an electric potential difference of $4.6\times10^3/\varepsilon = 5.09\times10^2$ Volts across a 1.4µm GaN layer in the A-sample of Ambacher et al. [14]. This large electric potential difference would require a large DC bias in the circuit to compensate for it, which should have been detected. If the band-gap limiting effect of Fiorentini *et al.* [18] is taken into account, screening charges of ~0.0005e/$a_o^2$ are required at both ends of the GaN layer in order to reduce the field down to 3.34Volts/1.4μm=0.024MV/cm. These screening charges are about twice larger in magnitude than the 2-D screening charge considered by Ambacher *et al.* to screen the difference $\sigma_{pol}$ (~0.00025 e/$a_o^2$) between $Al_{\sim0.275}Ga_{\sim0.725}N$ and GaN. Fourth, the square-well electron sheet model is unrealistic for an exponentially decaying screening charge distribution [32], which results from a balancing between drift and diffusion currents. In essence, the validity of the analysis of Ambacher *et al.* and the claim of the existence of $P_{BFV}$ in their paper [14] are questionable.

## 5. Conclusion

The polar [0001] Wurtzite GaN films are composed of dipolar $Ga^+$-$N^-$ bi-layers, which gives rise to an electrostatic potential difference across the film that tilts their energy bands. However, there is a self-regulated charge transfer across the film, which limits the tilt of bands, so that the local CBM at the $Ga^+$-terminated surface maintains approximate equality with the local VBM at the $N^-$-terminated surface. The first-principles calculation for the electric potential difference across the film and the spatial dependence of local CBM and VBM for 0.6-3.2 nm films confirm the self-regulated charge transfer effect. It is found that the saturation of the cross-film potential difference occurs for films thicker than ~4nm, so that this effect should be considered in most practical nitride multi-layer quantum wells for optoelectronic applications.

**Acknowledgement**

This work was supported by the National Science Council of Taiwan (contract number NSC 94-2120-M-110-002) and US DARPA (contract number N00014-00-1-0471).